\documentstyle{article}
\begin{document}

\renewcommand{\arraystretch}{1.0}
\renewcommand{\textfraction}{0.05}
\renewcommand{\topfraction}{0.9}
\renewcommand{\bottomfraction}{0.9}
\def\la{\;\raise0.3ex\hbox{$<$\kern-0.75em\raise-1.1ex\hbox{$\sim$}}\;}
\def\ga{\;\raise0.3ex\hbox{$>$\kern-0.75em\raise-1.1ex\hbox{$\sim$}}\;}
\def\lr{\;\raise0.3ex\hbox{$\rightarrow$\kern-1.0em\raise-1.1ex\hbox{$\leftarrow$}}\;}
\newcommand{\kB}{\mbox{$k_{\rm B}$}}           
\newcommand{\tn}{\mbox{$T_{{\rm c}n}$}}        
\newcommand{\tp}{\mbox{$T_{{\rm c}p}$}}        
\newcommand{\te}{\mbox{$T_{eff}$}}             
\newcommand{\dash}{\mbox{--}}                  
\newcommand{\ex}{\mbox{\rm e}}                 
\newcommand{\r}{\rule{0cm}{0.4cm}}
\newcommand{\hh}{\rule{0.5cm}{0cm}}
\newcommand{\hb}{\rule{0.4cm}{0cm}}
\newcommand{\rl}{\rule{0em}{1.8ex}}
\newcommand{\hhh}{\rule{1.4em}{0ex}}
\newcommand{\hhb}{\rule{1.2em}{0ex}}
\newcommand{\hhl}{\rule{2.4em}{0ex}}
\newcommand{\s}{$\;\;$}

\title{\bf Are Pulsars Bare Strange Stars?}

\author{R.X.~Xu, G.J.~Qiao, B.~Zhang\\
        CAS-PKU joint Beijing Astrophysical Center
        and Department of Astronomy,\\
\vspace{3mm}
        Peking University, Beijing 100781, China\\
\vspace{3mm}
{\it Presented in IAU Colloquium 177: Pulsar Astronomy and Beyond}\\
	{\it August 30 -- Setpember 3, 1999, Bonn, Germany}
}
\date{}

\maketitle

\begin{abstract}

It is believed that pulsars are neutron stars or strange
stars with crusts. However we suggest
here that pulsars may be {\it bare strange stars} (i.e.,
strange stars without crust).
Due to rapid rotation and strong emission, young strange
stars produced in supernova explosions should be bare
when they act as radio pulsars.
Because of strong magnetic field, {\it two} polar-crusts would
shield the polar caps of an accreting strange star.
Such a suggestion can be checked by further observations.

\end{abstract}

\vspace{5mm}

\noindent
Two greatest theories (i.e., gravity and quantum)
developed in this century resulted in the formation of the
theory of compact stars. One of such kind of stars observed
are pulsars that were discovered by radio astronomers in
1960s. More observations later in $X$-ray and $\gamma$-ray
bands confirmed the existence of pulsars in the nature:
a kind of celestial bodies with masses $\sim 1 M_{\odot}$,
radius $\sim 10$ km, and magnetic fields $\sim 10^{12}$
gausses. Such objects, named as pulsars, are popularly
thought to be neutron stars soon after the discoveries.

Owing to the development of the hadronic quark model in
1960s and 1970s, it is conjectured that strange quark
matter (SQM), composed of nearly equal numbers of up, down,
and strange quarks, may be an absolutely stable `hadron' of
strong interaction confined states. If this assumption is
true, strange star$^1$ is in the nature, which should be
a ground state of neutron star. Hence, radio pulsars
survived from supernovae might be strange stars$^1$.

It is addressed that a strange quark core is surrounded
by a normal matter crust in the conventional strange star
scenario for radio pulsar$^1$. However, bare strange stars
(BSS) can also {\it well} act as radio pulsars$^2$. If
pulsars are BSSs, 1, some problems in pulsar emission
mechanism can be settled$^{2,3}$, 2, strange stars and
neutron stars could {\it easily} be distinguished in
observation.

\vspace{5mm}
\noindent
{\bf Pulsars produced after supernovae: bare strange
stars?}~~
The process of phase transition from neutron matter to SQM
can cure the imperfection in the core-collapse supernova
paradigm$^4$. For progenitors with negligible effect of
rotation, additional neutrino emission by forming a strange
star can sufficiently enhance the power of neutrino
energy deposition behind the stalled shock, thus conduce
towards a successful explosion and its enough energy.
For
progenitors with rapidly rotating inner core, neutron
stars might be formed as semifinished products after
supernovae. Such a neutron star may be finally phase-converted
to a strange star when it spins down, and a $\gamma$-ray
burst may appear simultaneously$^4$.

There are two reasons for us to inevitably suggest that
a strange star as a radio pulsar should be bare. First,
the mass ejection rate from an envelope of a strange star
is very high soon after a supernova, therefore it is
natural to expect that the quark surface of a very young
strange star is nearly (or completely) bare$^5$. The
second one is the rapid rotation of newborn strange star.
{\it Only when the rotation of a strange star is slowed down
enough, would the accretion onto the surface be possible.}
The conditions for a possible accretion are
$
r_{\rm m}<r_{\rm c}<r_{\rm l}
$.
Substituting the radius of magnetosphere $r_{\rm m}$,
the co-rotation radius $r_{\rm c}$, and the radius of
light cylinder $r_{\rm l}$, we find
$P_1>0.7\stackrel{.}M_8^{-3/7}$ for typical pulsar values,
where $\stackrel{.}M_8$ is accretion rate in
$10^{-8}M_\odot \cdot {\rm yr}^{-1}$.
For solitary strange stars, $\stackrel{.}M_8$ is very
small because of very strong ejection around the star. If
$\stackrel{.}M_8=0.01$, only strange stars with periods
greater than 5s can accrete surrounding matter.

As $s$ quark is a little more massive than that of $u$ and
$d$ quarks, there are a few electrons in the chemical
equilibrium of SQM in order to keep the matter neutral.
The electromagnetic force participated in makes the
structure of strange quark matter more interesting and
attractive. It is found$^6$ that the number density of
electrons above quark surface
$
n_{\rm e} \sim {9.49 \times 10^{35} \over (1.2 z_{11} +
4)^3}
\; {\rm cm^{-3}},
$
and the electric field
$
E \sim {7.18 \times 10^{18} \over (1.2 z_{11} + 4)^2}
\; {\rm V\;cm^{-1}},
$
where $z$, in $10^{-11}$ cm, is a measured height above
the surface. These analytical results are valuable
if we are concerned about physical processes near the quark
surfaces of strange stars.

\vspace{5mm}
\noindent
{\bf An accreting pulsar: with two polar-crusts?}~~
As for strange stars in binaries, they could
be accretion powered $X$-ray sources with two
polar-crusts$^2$. In some cases, the intermittent phase
transition in polar-crust could result in a burst
process when one of the crusts is heavier than that the
Coulomb force can support. GRO J1744-28 is an ideal example
for studying such process. The observational phase-lag
in GRO J1744-28 may be a strong
evidence of the electric gap between strange quark
matter and polar-crust. The bursting can cause matter to
enough expand and to drag matter to latter phase. Such
phase-lag can be calculated to be $\sim 60^o$ if an
expanded crust is $\sim 1$ km.

\vspace{5mm}
\noindent
{\bf Conclusion}~~
We suggested that radio pulsars might be bare strange stars,
and that accreting strange stars have two polar-crusts.
Bare strange stars and neutron stars are {\it
distinguishable} in {\it observation} because their surfaces
differ in many ways, such as the electrodynamics, the
thermodynamics, and the composition (see, Xu, Qiao \&
Zhang in this peroceding). Detailed research on
this topic should be interesting and necessary.

\vspace{7mm}
We thank Mr. B.H. Hong, Dr. J.L. Han, and other members in our
pulsar group.
This work is supported by NSFC (No. 19803001,
N0.19910211260-570-A03), by the Climbing project of China, by
Doctoral Program Foundation of Institution of Higher Education
in China and by the Youth Foundation of PKU.

\begin{quote}

\verb"[1] Alcock, C., Farhi, E., Olinto, A. 1986, ApJ, 310, 261"\\
\verb"[2] Xu, R.X., Qiao, G.J. 1998, Chin Phys Lett, 15, 934"\\
\verb"[3] Xu, R.X., Qiao, G. J., Zhang, B. 1999, ApJ, 522, L109"\\
\verb"[4] Xu, R.X. et al. 1999, in K.S. Cheng (eds) Pacific Rim Conf.
on Stellar Astrophysics", \\Kluwer Publishers, in press\\
\verb"[5] Usov, V.V., 1998, Phys Rev Lett, 80, 230"\\
\verb"[6] Xu, R.X., Qiao, G.J. 1999, Chin Phys Lett, 16, 778"

\end{quote}

\end{document}